\documentclass[11pt]{article}
\usepackage{graphics}
\begin{document}
\title{Simpler proof of the theorem by Pusey, Barrett, and Rudolph on the reality of the quantum state}
\author{C. Moseley\\ Max-Planck-Institute for Meteorology, Hamburg, Germany
\\ \small email: christopher.moseley@mpimet.mpg.de}
\date{\today}
\maketitle

\begin{abstract}
The theorem of Pusey, Barrett, and Rudolph proves that different quantum states describe different physical realities. Their proof is based on the construction of entanglement measurement bases of two, and more than two qbits. In this note, I show that a two-qubit entanglement base is sufficient for a general proof.
\end{abstract}

\section{Introduction}
Recently, Pusey, Barrett, and Rudolph (hereafter abbreviated as PBR) proved an important theorem, showing that different quantum states represent physically different states of reality \cite{PBR}. Differently put, the theorem refutes the idea that one and the same physical reality can be consistently described by two different quantum states. EPR's proof is based on the idea that independent preparation devices can be constructed which can be set up to prepare a quantum system either in a state $|\psi\rangle$, or a different state $|\phi\rangle$. An ensemble of $n$ copies of this device ($n>2$) can then be used to prepare $n$ independent states $|C_{k_i}\rangle_i$, $i=1,\ldots,n$, with either $|C_{k_i}\rangle=|\psi\rangle$ or $|C_{k_i}\rangle=|\phi\rangle$, depending on the preparation $k_i\in\{1,2\}$ of device $i$. All these states are then destroyed in an $n$-partite entanglement measurement. Since each device can be prepared in two different ways, $2^n$ physically different situations can be construced. EPR construct entanglement measurements $M_n$ with $2^n$ different possible outcomes, in such a way that there is a bijective mapping of each of the $2^n$ possible device settings to one outcome which is measured with zero probability if this setting is chosen. If we assume that there is a non-zero probability of at least $\epsilon$ that a device produces a physical reality which is compatible with {\it both} states $|\psi\rangle$ and $|\phi\rangle$, it follows from the independent preparation of the states that the product state $|C_{k_1}\rangle_1\otimes\ldots\otimes |C_{k_n}\rangle_n$ would be compatible with {\it all} possible product states with a probability of at least $\epsilon^n$. However, this would mean that {\it all} possible outcomes of $M_n$ should be zero, which leads to a contradiction. Thus -- given quantum mechanics is correct -- we must conclude that two different quantum states can never be consistent with the same reality. Different proofs have been given, which lead to the same general conclusion \cite{Colbeck_Renner,Miller}. 

The tricky part of the proof by EPR is the construction of the basis of the entanglement measurement $M_n$. They show that the simplest case $n=2$ is sufficient to proove the theorem only for a subset of possible state pairs $(|\psi\rangle,|\phi\rangle)$ that are "not too close to each other"; more precisely, their scalar product must fulfill the inequality $|\langle\psi|\phi\rangle|\leq 1/\sqrt{2}$. For a general proof, a larger number  of independent preparation devices $n$ has to be used. Here, I present an argument that even in the general case, it is always possible to track the problem down to the case $n=2$, rendering the construction of entanglement measurement bases $M_n$ with $n>2$ unnecessary. 

In section 2, I recapitulate the proof for $n=2$, based on PBR's paper. In section 3, I present the alternative general proof, and section 4 compares both proofs.

\section{Proof for the case of two quantum states}

\begin{figure}
 \centering
 \includegraphics{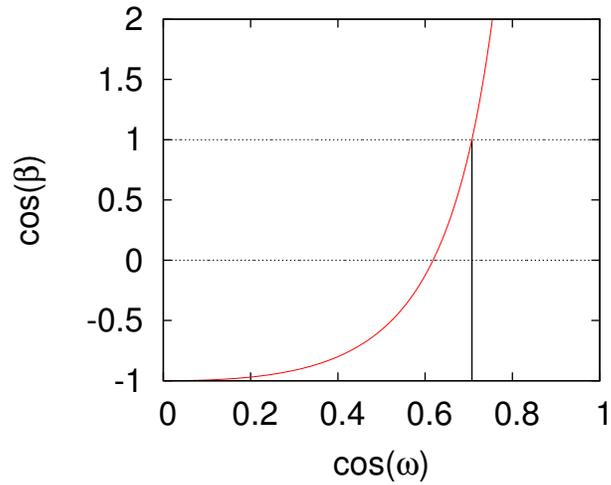}
 \caption{Value of $\cos(\beta)$ as a function of $\cos(\omega)$. The curve crosses the line $\cos(\beta)$=+1 at a value of $\cos(\omega)=\sqrt{2}/2\approx 0.7$ (marked by the vertical line), above which no real value for $\beta$ can be given.}
 \label{figure1}
\end{figure}

Two identical devices, labelled by $i=1,2$, can be prepared to either produce the quantum state $|\psi\rangle_i$ or $|\phi\rangle_i$. Independetly of the dimension if the Hilbert space $\mathcal{H}_i$ which $|\psi\rangle_i$ and $|\phi\rangle_i$ are defined in, a basis $\{ |0\rangle,|1\rangle \}$ can always be chosen such that (possibly after the multiplication of one of the states with a phase $e^{{\rm i}\phi}$, without any effect to the physical state of the system)
\begin{equation}
  \begin{array}{l@{\quad}l}
  |\psi\rangle_i = \cos\frac{\omega}2 |0\rangle_i + \sin\frac{\omega}2 |1\rangle_i \in  \mathcal{H}_i\\
  |\phi\rangle_i = \cos\frac{\omega}2 |0\rangle_i - \sin\frac{\omega}2 |1\rangle_i \in  
  \mathcal{H}_i\end{array} \; , i=0,1 \; , \mbox{ with } 0<\omega\leq \frac\pi 2 \; .
  \label{basis_qbit}
\end{equation}
Thus, the modulus of the scalar product the two is given as
\begin{equation}
 |\langle\psi|\phi\rangle_i|=\cos^2\frac\omega 2-\sin^2\frac\omega 2=\cos\omega \; .
 \label{def_omega}
\end{equation}
The crucial point here is, that regardless of the dimension of $\mathcal{H}_i$, the whole problem can be reduced to two two-dimensional Hilbert spaces, treating the quantum state generated by each device as a qbit. We define the four basis states of the product Hilbert space $\mathcal{H}_1\otimes\mathcal{H}_2$ by:
\begin{equation}
  \begin{array}{l@{\quad}l}
  |1\rangle := |0\rangle_1 \otimes |0\rangle_2 \\
  |2\rangle := |0\rangle_1 \otimes |1\rangle_2 \\
  |3\rangle := |1\rangle_1 \otimes |0\rangle_2 \\
  |4\rangle := |1\rangle_1 \otimes |1\rangle_2 \end{array}
\end{equation}
On this product space, we define an operator $C$, such that $C$ generates all four product states which can be jointly generated by the two devices:
\begin{equation}
  \begin{array}{l@{\quad}l}
  C|1\rangle = |\psi\rangle_1 \otimes |\psi\rangle_2 \\
  C|2\rangle = |\psi\rangle_1 \otimes |\phi\rangle_2 \\
  C|3\rangle = |\phi\rangle_1 \otimes |\psi\rangle_2 \\
  C|4\rangle = |\phi\rangle_1 \otimes |\phi\rangle_2 \end{array}
\end{equation}
In Matrix form, $C$ is given as
\begin{equation}
  C_{jk} := \langle j|C|k\rangle = \left( \begin{array}{cccc}
  \cos^2\frac\omega 2 & \cos^2\frac\omega 2 & \cos^2\frac\omega 2 & \cos^2\frac\omega 2 \\
  \cos\frac\omega 2 \sin\frac\omega 2 & -\cos\frac\omega 2 \sin\frac\omega 2 & \cos\frac\omega 2 \sin\frac\omega 2 & -\cos\frac\omega 2\sin\frac\omega 2 \\
  \cos\frac\omega 2 \sin\frac\omega 2 & \cos\frac\omega 2 \sin\frac\omega 2 & -\cos\frac\omega 2 \sin\frac\omega 2 & -\cos\frac\omega 2\sin\frac\omega 2 \\
  \sin^2\frac\omega 2 & -\sin^2\frac\omega 2 & -\sin^2\frac\omega 2 & \sin^2\frac\omega 2
 \end{array} \right) \; .
\end{equation}
A given product state $C|j\rangle$ with $j=1,\ldots,4$ is now subject to an entanglement measurement, with a measurement basis given by states $M^{-1}|k\rangle$, $k=1,\ldots,4$ generated by a unitary operator $M$ which in matrix form is given as
\begin{equation}
  M_{jk} := \langle j|M|k\rangle = \frac 12 \left( \begin{array}{cccc}
  e^{{\rm i}\alpha} &  e^{{\rm i}\beta} &  e^{{\rm i}\beta} &  e^{2{\rm i}\beta} \\
  e^{{\rm i}\alpha} & -e^{{\rm i}\beta} &  e^{{\rm i}\beta} & -e^{2{\rm i}\beta} \\
  e^{{\rm i}\alpha} &  e^{{\rm i}\beta} & -e^{{\rm i}\beta} & -e^{2{\rm i}\beta} \\
  e^{{\rm i}\alpha} & -e^{{\rm i}\beta} & -e^{{\rm i}\beta} &  e^{2{\rm i}\beta}
  \end{array} \right) \; ,
\end{equation}
with real numbers $\alpha$ and $\beta$. If the measurement returns the output $k$, the quantum system will be projected to the entagled state $M^{-1}|k\rangle$. Given that the system is prepared in the initial state $C|j\rangle$ by a corresponding setup of the two devices, the probability of recieving a measurement output $k$ is given as $|\langle k|MC|j\rangle|^2$. PBR show that under certain conditions it is possible to choose $\alpha$ and $\beta$ such that given the initial state $C|j\rangle$, the probability of measuring $j$ becomes zero. This is the case if
\begin{equation}
  \langle j|MC|j\rangle =
  \frac 12 {\rm e}^{{\rm i}\alpha} \cos^2\frac\omega 2 +
  {\rm e}^{{\rm i}\beta} \cos\frac\omega 2 \sin\frac\omega 2 +
  \frac 12 {\rm e}^{2{\rm i}\beta} \sin^2\frac\omega 2 \equiv 0 \; \mbox{ for all } j=1,\ldots,4 \; .
  \label{jMCj}
\end{equation}
This equation yields for $\beta$ the solution
\begin{equation}
  \cos\beta = \frac 14 \left( \tan^{-3}\frac\omega 2 - 
  4\tan^{-1}\frac\omega 2 - \tan\frac\omega 2 \right) \; .
\end{equation}
Using the identity
\begin{equation}
  \tan^2\frac\omega 2 = \frac{1-\cos\omega}{1+\cos\omega}
  \label{tan_cos_identity}
\end{equation}
this equation can also be written as
\begin{equation}
  \cos\beta=\frac{\cos^2\omega+\cos\omega-1}{
  \left(1-\cos\omega\right)\sqrt{1-\cos^2\omega}} \; .
  \label{cosbcosw}
\end{equation}
The relationship between $\beta$ and $\omega$ in eq. (\ref{cosbcosw}) is plotted in Fig. \ref{figure1}, showing that $-1\leq\cos\beta\leq 1$ is only fulfilled in the range $\cos\omega\le \sqrt{2}/2$. If $\cos\omega>\sqrt{2}/2$, $\cos\beta$ becomes larger than $1$, meaning that no real value for $\beta$ can be given in order to fulfill eq. (\ref{jMCj}).

The assumption as already mentioned in the Introduction, that both devices independently produce a physical state which is compatible with $|\psi_i\rangle$ and $|\phi_i\rangle$ for $i=1,2$ at the same time with a non-zero probability of at least $\epsilon^2$, leads to the desired contradiction with eq. (\ref{jMCj}): In these cases, the product state would be compatible with all $C|j\rangle, \; j=1,\ldots,4$, meaning that all four possible measurement outcomes would get zero probability. Note, however, that this argument fails if $|\langle\psi|\phi\rangle|>1/\sqrt{2}$, and is thus not sufficient for a general proof.

\section{Alternative proof of the general case}

We now assume arbitrary states $|\psi\rangle$ and $|\phi\rangle$, without any restriction to $\omega$. We consider an ensemble of $n$ independent copies of the preparation device, with $n$ being an even number such that the devices can be divided into two groups, say, group 1 and 2, each consisting of $n/2$ devices. We denote the devices in group 1 by $i=1,\ldots,n/2$, and in group 2 by $i=n/2+1,\ldots,n$. All devices in group 1 can be prepared to generate a product state $|C_1\rangle_1\otimes\ldots\otimes |C_1\rangle_{n/2}=|C_1\rangle^{\otimes(n/2)}$, with either $|C_1\rangle=|\psi\rangle$, or $|C_1\rangle=|\phi\rangle$. Respectively, all deviced in group 2 can be prepared to generate a state $|C_2\rangle_{n/2+1}\otimes\ldots\otimes |C_2\rangle_{n}=|C_2\rangle^{\otimes(n/2)}$. We can now define:
\begin{eqnarray}
  |\Psi\rangle_1 &:=& |\psi\rangle_1\otimes\ldots\otimes |\psi\rangle_{n/2} 
    \in \bigotimes_{i=1}^{n/2} \mathcal{H}_i\\
  |\Phi\rangle_1 &:=& |\phi\rangle_1\otimes\ldots\otimes |\phi\rangle_{n/2}
    \in \bigotimes_{i=1}^{n/2} \mathcal{H}_i\\
  |\Psi\rangle_2 &:=& |\psi\rangle_{n/2+1}\otimes\ldots\otimes |\psi\rangle_{n}
    \in \bigotimes_{i=n/2+1}^n \mathcal{H}_i\\
  |\Phi\rangle_2 &:=& |\phi\rangle_{n/2+1}\otimes\ldots\otimes |\phi\rangle_{n} 
    \in \bigotimes_{i=n/2+1}^n \mathcal{H}_i\\
\end{eqnarray}
Using eq. (\ref{def_omega}), the scalar product of these product states fulfills
\begin{equation}
  \langle\Psi|\Phi\rangle_1 = \langle\Psi|\Phi\rangle_2 =
  |\langle\psi|\phi\rangle_i|^{\frac n2} = \cos^{\frac n2} \omega \equiv \cos\Omega
  \label{def_gross_omega}
\end{equation}
where we define $\Omega$ with $0<\Omega\leq \frac\pi 2$, in analogy to $\omega$ in eq. (\ref{def_omega}). Since $\cos\omega<1$, it is always possible to choose $n$ large enough such that $\cos\Omega\le \sqrt{2}/2$. If we now apply the PBR assumption as stated in the Introduction, that there is at least the probability $\epsilon^n$ that all states $|C_1\rangle_i$, $i=1,\ldots,n/2$ in group 1 and $|C_2\rangle_i$, $i=n/2+1,\ldots,n$ in group 2 are each compatible with both $|\psi\rangle_i$ and $|\phi\rangle_i$, this would imply that with non-zero probability, the product state $|C_1\rangle^{\otimes(n/2)}$ were compatible with both $|\Psi\rangle_1$ and $|\Phi\rangle_1$, and at the same time, $|C_2\rangle^{\otimes(n/2)}$ were compatible with both $|\Psi\rangle_2$ and $|\Phi\rangle_2$. Thus, we can to perform the replacements $\Psi\rightarrow\psi$,  $\Phi\rightarrow\phi$, and $\Omega\rightarrow\omega$ and repeat the two-qbit proof given in Section 2 for the product states $|\Psi\rangle$ and $|\Phi\rangle$. This is possible, since in eq. (\ref{basis_qbit}) there is no restriction to the Hilbert space in which the two states are defined in. Using these replacements, we yield a general proof of the PBR theorem, by reducing it to the two-qbit case.

\section{Comparison to PBR's proof and Conclusion}

\begin{figure}
 \centering
 \includegraphics{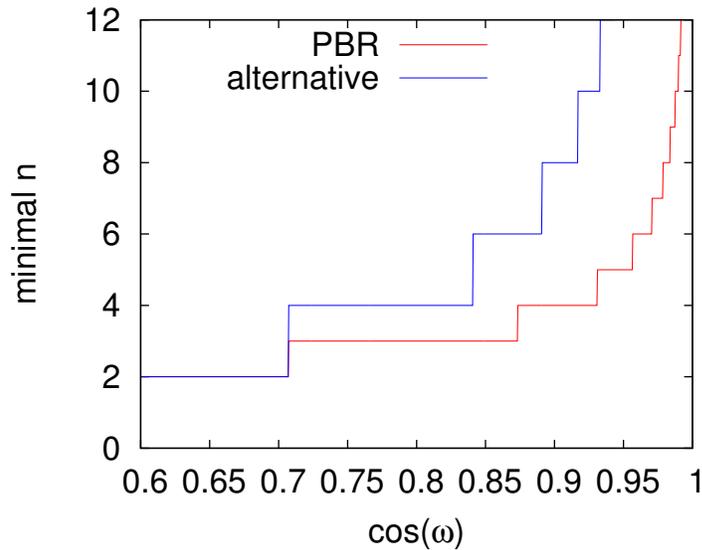}
 \caption{Minimum number $n$ of independently prepared quantum states necessary to prove the PBR theorem for given value of $\cos(\omega)$. Red color: Proof given in PBR's paper. Blue color: Alternative proof presented in this note.}
 \label{figure2}
\end{figure}

PBR prove the general case by constructing a hierarchy of multi-partite entanglement measurements, where the number of independently prepared quantum states $n$ must be chosen such that
\begin{equation}
  \tan\frac\omega 2 \geq 2^{1/n}-1 \; ,
\end{equation}
which is equivalent to
\begin{equation}
  n \geq \frac{\ln 2}{\ln\left( 1+\sqrt{\frac{1-\cos\omega}{1+\cos\omega}}\right)} \; ,
  \label{n_PBR}
\end{equation}
where again the identity eq. (\ref{tan_cos_identity}) has been used. The proof presented here is identical to PBR's for the case $n=2$, i.e. $\tan\frac\omega 2\geq \sqrt{2}-1$, being equivalent to
\begin{equation}
\cos\omega\leq \frac{\sqrt 2}2 \; .
\end{equation}
Otherwise, according to eq. (\ref{def_gross_omega}), $n$ has to be chosen such that 
\begin{equation}
 \cos\Omega=\cos^{\frac n2} \omega \leq \frac{\sqrt 2}2 \; ,
\end{equation}
leading to the condition
\begin{equation}
 n\geq -\frac 12 \, \frac{\ln 2}{\ln\cos\omega} \; ,
  \label{n_alternative}
\end{equation}
with $n$ being an even number, for the choice of $n$. The minimal number of $n$ to prove the theorm with a given value of $\cos\omega$, is give in Fig. \ref{figure2} for both the PBR prove and the alternative proof, as given by eqs. (\ref{n_PBR}) and (\ref{n_alternative}), respectively. The graph shows, that a larger number of indepentend preparation devices is necessary for the alternative proof, if $\cos\omega>\sqrt{2}/2\approx 0.7$. Therefore, we could state that BPR's original proof is more "efficient". On the other hand, the proof presented here works without the construction of a non-trivial multi-partite entanglement measurement basis for more than two qbits, but rests on a simple argument in the case where BPR's proof does not work with two qbits only. It shows that the general case follows immediately from the two qbit case. It is therefore simpler than the proof given by PBR.

%------------------------------------------------------------------------------------------

\end{document}